\documentstyle[aas2pp4]{article}
\newcommand{\be}{\begin{equation}}
\newcommand{\ba}{\begin{eqnarray}}
\newcommand{\ee}{\end{equation}}
\newcommand{\ea}{\end{eqnarray}}
\newcommand{\sol}{M_{\sun}}

\newcommand{\etal}{{\it et al. }}

\lefthead{Nakamura et al. }
\righthead{The Minimum Total Mass of MACHOs and Halo Models of the Galaxy}

\begin{document}

\thispagestyle{empty}
{\baselineskip0pt
\leftline{\large\baselineskip16pt\sl\vbox to0pt{\hbox{Yukawa Institute}
               \hbox{Kyoto University}\vss}}
\rightline{\large\baselineskip16pt\rm\vbox to20pt{\hbox{YITP-96-49}
               \hbox{KUNS-1415}
\vss}}%

\title{The Minimum Total Mass of MACHOs and Halo Models of the Galaxy}

\author{T. Nakamura}
\affil{Yukawa Institute for Theoretical Physics, Kyoto University,
Kyoto 606}

\and
\author{Y. Kan-ya and R. Nishi}
\affil{Department of Physics , Kyoto University , Kyoto 606}
\received{ 4 June   1996}
\accepted{ 11 October 1996}

\begin{abstract}
If the density distribution $\rho (r)$ of MACHOs is spherically
symmetric with respect to the Galactic center, it is
shown that the minimal total mass $M_{min}^{{\rm MACHO}}$ of the MACHOs is
 $1.7\times 10^{10}\sol \tau_{-6.7}^{{\rm LMC}}$ where $
\tau_{-6.7}^{{\rm LMC}}$ is the optical depth ($\tau^{{\rm LMC}}$)
toward the Large Magellanic
Cloud (LMC)  in the unit of $2\times 10^{-7}$. If $\rho (r)$ is a decreasing function of $r$, it is
proved that $M_{min}^{{\rm MACHO}}$ is
 $5.6\times 10^{10}\sol \tau_{-6.7}^{{\rm LMC}}$.   
 Several spherical and axially symmetric halo models of the Galaxy
 with a few free parameters are also considered.  It is found that
$M_{min}^{{\rm MACHO}}$ ranges from $  5.6\times 10^{10}\sol \tau_{-6.7}^{{\rm LMC}}$ to
$ \sim 3 \times 10^{11}\sol \tau_{-6.7}^{{\rm LMC}}$. For
general case, the minimal column density 
$\Sigma_{min}^{{\rm MACHO}}$ of MACHOs is obtained as 
$\Sigma_{min}^{{\rm MACHO}} =25 \sol {\rm pc}^{-2}\tau_{-6.7}^{{\rm LMC}}.
 $ If the clump of MACHOs exist only halfway between LMC and
the sun, $ M_{min}^{{\rm MACHO}}$ is $1.5\times 10^9\sol$.  This shows that
the total mass of MACHOs is smaller than $5 \times 10^{10}\sol $
, i.e. $\sim$ 10\% of the mass of the halo inside  LMC, either if the density distribution of MACHOs is
unusual or $ \tau^{{\rm LMC}}\ll 2\times 10^{-7}$.

(11 October 1996, Accepted for publication in Apj. Letters.)
\end{abstract}

\keywords{dark matter --- Galaxy: halo --- Galaxy: structure --- 
gravitational lensing}

\section{Introduction}
  Recent second year analysis of microlensing events toward the Large
Magellanic Cloud (LMC) by the MACHO
collaboration suggests that the optical depth $\tau^{{\rm LMC}}$ is $\sim 2\times
10^{-7}$ and the fraction $f$  of MACHOs is $\sim 0.4$ with the typical
mass $\sim 0.34 \sol $ in the standard spherical flat rotation halo 
model (\cite{benn96}).  The estimated mass of MACHOs is just the mass of 
red dwarfs. However the contribution of the  halo red dwarfs to MACHO 
events should be small  since  the observed density of the halo red dwarfs 
is too low (\cite{bahc94,graf96a,graf96b}).

As for the white dwarf Galaxy halo, Charlot \& Silk (1995) combined
population synthesis models with constraints from deep galaxy surveys
and showed that only a small fraction ($\leq 10\%$) of the dark mass in
the present-day galaxy halo could be in the form of white dwarf remnants
(WDR) of intermediate-mass stars.  Adams \& Laughlin (1996) recently
argued the implications of white dwarf Galactic halos. From the current
limits on the density of red dwarfs (\cite{bahc94,graf96a,graf96b} ) and
the galactic metallicity, the IMF must be sharply peaked about a
characteristic mass scale $M_c\sim 2.3\sol$ . They concluded that the
mass fraction of WDRs in the halo is likely to be less than 25\% since
only a fraction of the initial mass of a star is incorporated into WDRs.

The spatial density distribution function of MACHOs which caused
microlensing events is not known in spite of many arguments on the mass
distribution of halo dark matter
(\cite{pacz86,grie91,sack93,frie94,sahu94,alco95,gate96,turn96,evan96,kany96}).
Only the possible value of the optical depth $\tau^{{\rm LMC}}
$(\cite{benn96}) is known, the fraction of MACHOs in the halo depending
on the spatial density distribution function of MACHOs. In this
situation it is important to check the relation between the total mass
$M^{{\rm MACHO}}$ of MACHOs to the density distribution function of
MACHOs for the given optical depth $\tau^{{\rm LMC}}$. The results of
such a study will be useful in the arguments on the fraction of MACHOs
in the halo and what MACHOs are.  

In this paper we study $M^{{\rm MACHO}}$ for various density
distribution functions of MACHOs and discuss the minimal total mass of the
MACHO halo.  Gates, Gyuk, and Turner (1996) and Turner, Gates, and Gyuk
(1996) also discussed the mass of the MACHO halo for various galaxy models. While they separately
added unidentified dark thick disk components, we discuss the total mass
responsible for microlensing.  In \S 2 we obtain the minimal total mass
$M_{min}^{{\rm MACHO}}$ of MACHOs for spherically symmetric density
distributions. In \S 3 we discuss $M_{min}^{{\rm MACHO}}$ for various
axially symmetric density distributions. \S 4 will be devoted to
discussions.

\section{Spherically Symmetric Halo Models and the Minimal Total Mass
of  MACHOs} 
 We assume that the density distribution function $\rho (r)$ of MACHOs 
is  a function of the galactocentric radius $r$. The optical
depth $\tau^{{\rm LMC}}$ toward LMC is given by (\cite{pacz86,grie91}) 
\be
\tau^{{\rm LMC}}=\frac{4\pi G}{c^2}\int_0^{D_s}x(1-\frac{x}{D_s})\rho (r)dx, 
\ee     
\be
r^2=R_0^2-2R_0\eta x +x^2, 
\ee
and
\be
\eta=\cos b \cos l ,
\ee
where $D_s, l,b$ and $ R_0$ are the distance to LMC (50kpc),
 the galactic longitude
and latitude of LMC and the galactocentric radius of the sun (8.5kpc),
respectively. In Eq. (1) we assumed the threshold $u_T$=1 for simplicity.

Equation (1) is rewritten as
\be
\tau^{{\rm LMC}}=\int f(x) dm, 
\ee 
\be
f(x)=\frac{ G}{c^2}\frac{x(1-\frac{x}{D_s})}{r^2\frac{dr}{dx}}, 
\ee 
and 
\be
dm=4\pi r^2\rho (r)\frac{dr}{dx}dx. 
\ee 
For  LMC,  $f(x)$ is infinite  at
$x=x_c\equiv R_0\eta=0.153R_0$ so that  the minimal total mass 
$M_{min}^{{\rm MACHO}}$ of MACHOs is zero for any given $\tau^{{\rm LMC}}
$  if MACHOs are distributed in an infinitesimally thin shell at $r=r_c\equiv
\sqrt{1-\eta^ 2}R_0$. However this is wrong.  Since the angular size of LMC is 
$\sim 10\arcdeg \times 10\arcdeg $, $M^{{\rm MACHO}} $ is minimized if MACHOs are
distributed in a shell at $r=r_c$ with width $d$ given by
\be
d = \frac {10\pi }{180}R_0 \eta =227{\rm pc} .
\ee
It is easy to show that $M_{min}^{{\rm MACHO}}$ is given by
\be
M_{min}^{{\rm MACHO}}= \frac{c^2\tau^{{\rm LMC}}}{3G} \frac {(r_c+d)^3-r_c^3}
{2\sqrt{2r_cd+d^2}(R_0\eta-\frac{R_0^2\eta^2+6r_cd+3d^2}{3D_s})},
\ee
\be
 =1.7\times 10^{10}\sol \tau_{-6.7}^{{\rm LMC}}, 
\ee
where $\tau_{-6.7}^{{\rm LMC}}$  is $\tau^{{\rm LMC}}$ in the unit of $2\times 10^{-7}$.     
This shows that in principle $M^{{\rm MACHO}} $ can be only $\sim 3\%$ of
the total mass of the halo inside LMC. However the density distribution
function of MACHOs in this case is very peculiar so that we calculate $M^{{\rm MACHO}} $ for more
realistic $\rho (r)$ to know more realistic 
$M_{min}^{{\rm MACHO}}$. We  consider two models;
 
1)Polytropic Model

$\rho (r)$ is given by polytrope of index N and the
radius $R_p$.

2) $\alpha$ Model

$\rho (r)$ is given by
\be
\rho (r)=\frac {\rho_0} {(1+\frac{r^2}{R_a^2})^\alpha}. 
\ee
This model is similar to the beta model of the cluster of galaxies with 
core radius $R_a$.

In figure 1 we show $M^{{\rm MACHO}}$ as a function of $R_p$ for several
polytropic indices N. $M_{min}^{{\rm MACHO}}$ ranges from $5.6\times
 10^{10}\sol \tau_{-6.7}^{{\rm LMC}}$ for N=0 to $7.8\times
10^{10}\sol \tau_{-6.7}^{{\rm LMC}} $ for N=3. For N=4
and 4.5,  $M^{{\rm MACHO}}$ is greater than $1.0 \times 10^{11}\sol 
\tau_{-6.7}^{{\rm LMC}}$ and the minimum does not exist 
for $R_p < D_s$.  
Under the assumption that $\rho (r)$ is 
a  decreasing function of $r$,
it is shown in the Appendix that $M^{{\rm MACHO}}$ is
 minimized when $\rho (r)$ is constant. Therefore $M_{min}^{{\rm MACHO}}$ 
is $5.6\times 10^{10}\sol \tau_{-6.7}^{{\rm LMC}}$ if
 $\rho (r)$ is a decreasing function. In figure 2 we show the total
mass of MACHOs inside LMC in $\alpha$ models as a function of $R_a$
for several values of $\alpha$.  $M_{min}^{{\rm MACHO}}$ ranges from
$1.3 \times 10^{11}\sol \tau_{-6.7}^{{\rm LMC}}$ for $\alpha =1.5$ to
$8.7\times 10^{10}\sol \tau_{-6.7}^{{\rm LMC}}$ for $\alpha
=6$. For large $\alpha$,  $M_{min}^{{\rm MACHO}}$ does not change so much and
it converges although the value of $R_a$ at the minimum 
increases . This behavior can be understood analytically using the
asymptotic expression of gamma functions.

\section{Axially Symmetric Halo Models and the Minimal Total Mass
of the MACHOs} 
 There are several suggestions that the Galactic halo is not spherically
symmetric(\cite{aars78,agui90,binn94}) so that we study here axially symmetric halo models and
calculate $M^{{\rm MACHO}}$. We consider two models;

1)  Exponential Disk Model 

The axially symmetric density distribution function  $\rho(R,Z)$ in 
cylindrical coordinates is given by
\be
\rho(R,Z)= \rho_0 \exp (-\frac{R}{R_d}-\frac{\mid Z \mid}{Z_d}), 
\ee
where  $R_d$ and $Z_h$ are scale heights.
 
2) Elliptical Model

 $\rho(R,Z)$ is given by
\be
\rho(R,Z)= \frac {\rho_0}{(1+\frac{R^2}{a^2}+\frac{Z^2}{c^2})^\alpha }, 
\ee
where $a$ and $c$ describe the ellipticity of the equidensity surface.
This is an axially symmetric version of the $\alpha$ model.

In figure 3 we show $M^{{\rm MACHO}}$ as a function of $R_d$ in exponential
 disk models for various aspect ratios ($Z_h/R_d$) of the equidensity surface. 
We see  $M_{min}^{{\rm MACHO}}$ is $\sim 1.0\times  10^{11}\sol
\tau_{-6.7}^{{\rm LMC}}$ for $0.5< Z_h/R_d <1.0$ and it 
increases with the decrease of $Z_h/R_d$ for $Z_h/R_d <0.5$.

In figure 4a and 4b we show $M^{{\rm MACHO}}$ in elliptical models as a
function of $a$ and the aspect ratio $c/a$ for $\alpha =2.5$ and
$\alpha =6.0$, respectively.  $M_{min}^{{\rm MACHO}}$ is $8.9\times  10^{10}\sol\tau_{-6.7}^{{\rm LMC}}$ at $a$=10kpc and $c$=6kpc for  $\alpha
=2.5$ and is $7.08\times  10^{10}\sol \tau_{-6.7}^{{\rm LMC}}$
at $a$=22kpc and $c$=13.2kpc  for $\alpha =6.0$. For large $\alpha$
, $M_{min}^{{\rm MACHO}}$ converges, similarly to the $\alpha$ models.

\section{Discussions}

A deep north Galactic pole proper motion survey(\cite{maje96})
suggests that the halo is not dynamically mixed but contains a
significant fraction of stars with membership in correlated stellar streams.
If MACHOs are also dynamically unmixed, it is possible that the
density distribution function is neither spherically nor axially symmetric but 
completely inhomogeneous. In such a case what we can say from the
microlensing events toward LMC is the minimal column density 
$\Sigma_{min}^{{\rm MACHO}}$ of MACHOs. 
Since in Equation (1), $x(1-x/D_s) < D_s/4$,
   $\Sigma_{min}^{{\rm MACHO}}$ is given by 
\be
\Sigma_{min}^{{\rm MACHO}} =25 \sol {\rm pc}^{-2}
\tau_{-6.7}^{{\rm LMC}}.
\ee 
Similar to Equation (7), the linear size of the clump of MACHOs should 
be larger than 174pc$(x/$kpc) where $x$ is the distance to the clump 
of MACHOs .
For $x=D_s/2,   M_{min}^{{\rm MACHO}}$ is $1.5\times 10^9\sol$. If this 
is the case, the optical depth toward the Small Magellanic Cloud will
be quite different and the inhomogeneity of the density distribution of MACHOs 
can be checked.

In conclusion, it is shown that the total mass of MACHOs becomes smaller than $5 \times 10^{10}\sol $
, i.e. $\sim$ 10\% of the mass of the halo inside the LMC, either if the density distribution of MACHOs is
unusual or $ \tau^{{\rm LMC}}\ll 2\times 10^{-7} $.

\acknowledgments
 We like to thank Hayward for checking English.
 This work was supported by
Grant-in-Aid of Scientific Research of the Ministry of Education,
Culture, and Sports, No.07640399.

\appendix
 {\bf Appendix}

{\bf The proof that $M^{{\rm MACHO}}$ is minimized if the spherically
symmetric density distribution function $\rho (r)$ is a constant.}

 We first fix the density at $r=r_c$. Since the matter inside  
$r_c$ does not contribute to $\tau $, $M^{{\rm MACHO}}$ is minimized if
$\rho (r)$ is constant for $r<r_c$. For  $r\ge r_c$ from Equation (4) 
it is easy to observe that $ d\tau /dm$ is a decreasing function of $r$
 irrespective of $\rho (r)$. Now for fixed $m$ (i.e. the same mass but 
different $r$ depending on the density distribution function),
  $r$ is smallest 
and $ d\tau /dm$ is  largest if $\rho (r)=\rho(r_c)$
=constant. This means that $\tau$ is largest for constant 
density distribution for fixed total mass $M^{{\rm MACHO}}$ . 
Inversely for fixed $\tau$,  $M^{{\rm MACHO}}$  is minimized
for the constant density distribution. By varying the value $\rho (r_c)$
, $M_{min}^{{\rm MACHO}}$ is obtained.

%
%

\newpage

\figcaption{ The total mass of MACHOs for the polytropic model in the unit 
of $\sol$. N is the
polytropic index and $R_p$ is the radius of the density distribution of MACHOs.}
\figcaption{ The total mass of MACHOs for the $\alpha$ model  in the unit 
of $\sol$. The density
distribution is given by $\rho (r)=\frac {\rho_0} {(1+\frac{r^2}{R_a^2})^\alpha}.  $}
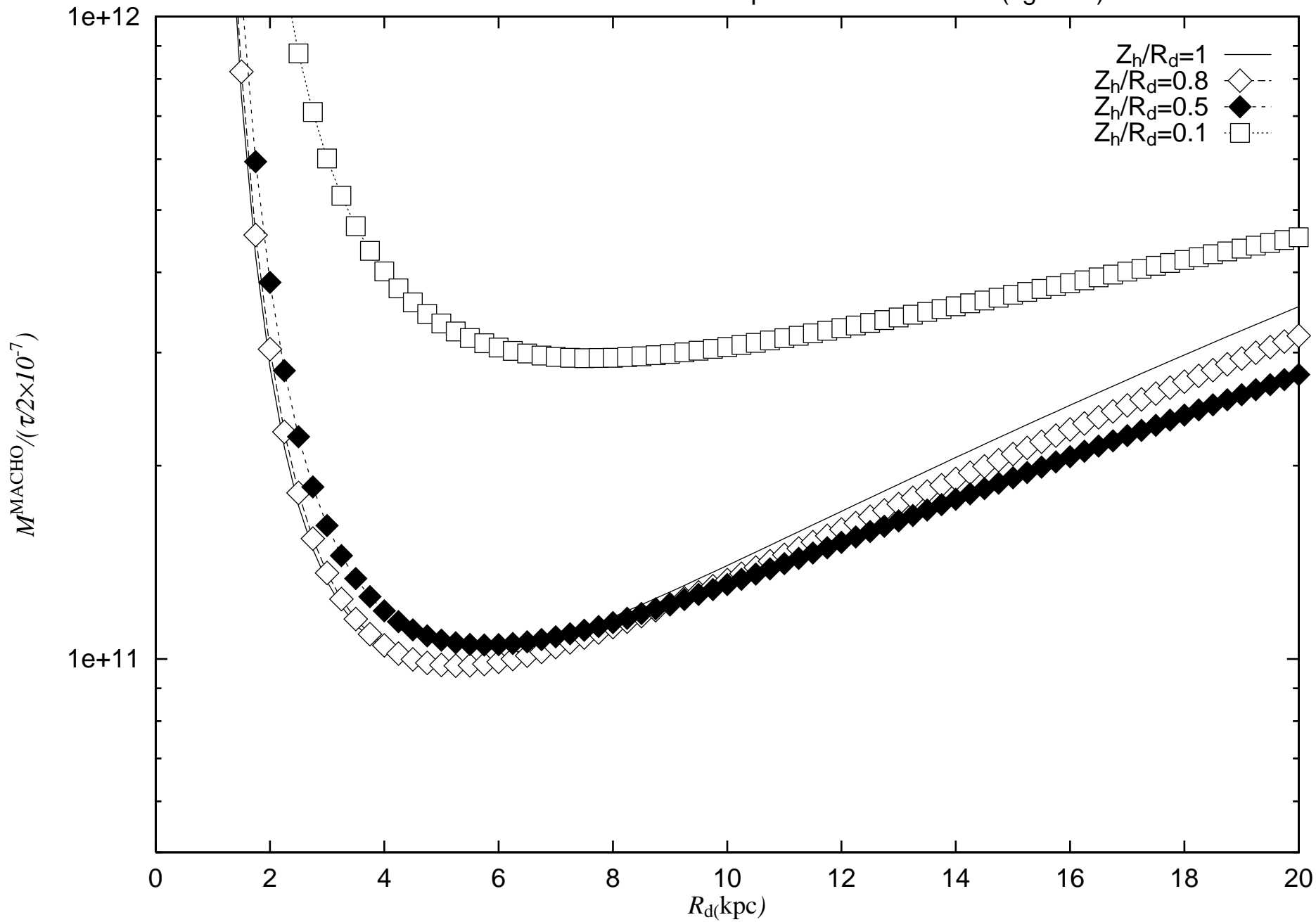
\figcaption{ The total mass of MACHOs for the exponential disk model  in the unit 
of $\sol$. The
density distribution is given by $\rho(R,Z)= \rho_0 \exp (-\frac{R}{R_d}-\frac{\mid Z \mid}{Z_d}).$ 
  }
\figcaption{ The total mass of MACHOs for the elliptic model  in the unit 
of $\sol$ . The density
distribution is given by $\rho(R,Z)= \frac
{\rho_0}{(1+\frac{R^2}{a^2}+\frac{Z^2}{c^2})^\alpha }. $ a) for
$\alpha =2.5$ and b) for $\alpha =6.0$}

\end{document}